\documentclass[12pt]{iopart}

\usepackage{setstack}
\usepackage{iopams}
\usepackage{amstext}
\usepackage{pxfonts}
\usepackage{graphicx}
\usepackage{eucal}
\usepackage{mathrsfs}
\usepackage{pifont}

\newcommand{\brk}[1]{\left( #1 \right)}

\newcommand{\BRK}[1]{\left\{ #1 \right\}}

\newcommand{\ao}{\bar{\mathfrak{a}}}
\newcommand{\bo}{\bar{\mathfrak{b}}}
\renewcommand{\a}{\mathfrak{a}}
\renewcommand{\b}{\mathfrak{b}}

\newcommand{\of}{\overline{f}}
\newcommand{\beq}{\begin{equation}}
\newcommand{\eeq}{\end{equation}}

\newcommand{\eqref}[1]{(\ref{#1})}

\newcommand{\En}{\mathcal{E}}
\newcommand{\EB}{\En_B}
\newcommand{\ES}{\En_S}

\begin{document}

\title[Metrically discontinuous non-Euclidean plates]{Pattern selection and multiscale behavior in metrically discontinuous non-Euclidean plates}
\author{Michael Moshe$^1$, 
Eran Sharon$^1$,
Raz Kupferman$^2$}

\address{$^1$ Racah Institute of Physics, The Hebrew University, Jerusalem 91904 Israel}
\address{$^2$ Einstein Institute of Mathematics, The Hebrew University, Jerusalem 91904 Israel}

\ead{\mailto{michael.moshe@mail.huji.ac.il}, \mailto{erans@vms.huji.ac.il}, \mailto{raz@math.huji.ac.il}}

\begin{abstract}
We study equilibrium configurations of non-Euclidean plates, in which the reference metric is uniaxially periodic. This work is motivated by recent experiments on thin sheets of composite thermally responsive gels \cite{WMGTNSK12}. Such sheets bend perpendicularly to the periodic axis in order to alleviate the metric discrepancy. For abruptly varying metrics, we identify multiple scaling regimes with different power law dependences of the elastic energy $\En$ and the axial curvature $\kappa$ on the sheet's thickness $h$. In the $h\to0$ limit the equilibrium configuration tends to an isometric embedding of the reference metric, and $\En\sim h^2$.
Two intermediate asymptotic regimes emerge in between the buckling threshold and the $h\to0$ limit, in which the energy scales either like $h^{4/5}$ or like $h^{2/3}$. We believe that this system exemplifies a much more general phenomenon, in which the thickness of the sheet induces a cutoff length scale below which finer structures of the metric cannot be observed. 
When the reference metric consists of several separated length scales, a decrease of the sheet's thickness results in a sequence of conformational changes, as finer properties of the reference metric are  revealed.
\end{abstract}


\submitto{Nonlinearity}


\section{Introduction}

Thin elastic sheets are being used extensively in both natural and manufactured structures to form elaborate three-dimensional (3D) configurations. Complex configurations are often a result of the material's  internal structure, rather than due to the application of external forces or constraints. In organic tissues, for example,  pattern formation is often a consequence of the tissue's nonuniform growth, or a response to differential swelling. 

The connection between  growth, or swelling, and the resultant 3D configurations is highly non-trivial. Different formulations of effective elastic theories of growing sheets have been proposed \cite{ESK08,BG05,Yav10}. In most approaches, gradients of growth (or swelling) across the thickness of the sheet induce a non-zero \emph{reference curvature} tensor, whereas  lateral gradients of growth endow the sheet with a 2D (generally non-Euclidean) \emph{reference metric} \cite{Wan67,Kro81,ESK08,KES07,San09,GV11}. 

Every configuration adopted by the sheet has associated metric and  curvature tensors (also known as first and second \emph{fundamental forms} \cite{DOC76}). Deviations of the actual metric from the reference metric cost \emph{stretching energy}, whereas deviations of the 
actual curvature from the reference curvature cost \emph{bending energy}. Shape selection is determined by an interplay between stretching and bending energies; the selected configuration  (i.e. the actual curvature and metric tensors) is the one that minimizes the \emph{total elastic energy}. 

Recent developments of experimental techniques \cite{KES07,KHHS12,WMGTNSK12} have promoted the joint theoretical and experimental study of pattern formation in numerous systems. In particular, various biologically-inspired elastic problems have been studied recently using  formulations of the type described above. Metric-driven shape selection has been studied in the context of leaves \cite{LM09}, fungi \cite{DB08}, petals \cite{MP06}, and various ribbon-like configurations \cite{San09,ESK11}. Curvature-driven shaping was shown to dominate shape transitions in flowers \cite{FSDM05}, pines \cite{DVR97}, and seed pod opening \cite{AESK11}. A similar geometric approach gave  rise also to reduced theories for non-Euclidean rod-like structures \cite{KS12}, and to a quantitative study of shape transitions in awns \cite{EZBF07} and isolated cells \cite{ATKDFRE11,AAESK12}. 

In all the above theoretical studies, the reference metric and curvature tensors were assumed to be smoothly varying relative to the lateral dimensions of the sheet. In reality, biological materials are often highly heterogeneous, both in their elastic moduli and in their growth/swelling profiles. The smooth reference tensorial fields that were used in the models have therefore to be viewed as \emph{homogenizations} of  fluctuating fields.  Homogenization was successfully applied,  for example, in \cite{KHBSH12}, were a ``digital" metric, of shrinking dots embedded within a less metrically responsive matrix was replaced by a smooth effective reference metric. Similarly, the fibrous structure of a pod valve \cite{AESK11} and a cell wall \cite{AAESK12} were successfully replaced by smooth effective reference metrics and curvatures. 

Naively, one would believe that the homogenization of the reference tensors is applicable when the thickness of the sheet is large compared to the scale of metric fluctuations.  Yet, the precise interplay between small scale metric fluctuations and the thickness of the sheet remains to be investigated, both experimentally and theoretically. In particular, it is not clear what to expect when the reference metric consists of several, well separated, length scales. 
Such questions are of major importance for the modeling of biological tissues and for the modeling of responsive composite materials, and are the main motivation for the present work.

This paper studies a common biological structure---a monolayer of a fibrous tissue. 
Fibrous tissues are ubiquitous, and appear in various sizes, ranging from cell walls of plants \cite{FZ55} and bacteria, and up to macrostructures. Fibrous tissues typically consist of aligned fibers embedded within a matrix. In some cases, such as in sclerenchymal tissue 
(rigid supporting plant tissue) \cite{FW72}, the swelling/growth properties of the matrix and  the fibers are different. As a result, changes in environmental conditions, such as humidity, or the active growth of the organ, lead to the buildup of internal stresses that are followed by shape transitions. 

It was recently shown \cite{RM09,AESK11} that sheets made of \emph{two} super-imposed fibrous layers behave like shells (i.e., have a non-zero reference curvature), and undergo various shape transitions.  
More recent experiments studied the response of \emph{monolayers}  of fibrous tissues to environmental changes. From a theoretical point of view, such structures form a new kind of non-Euclidean plate, with a reference metric that has a striped structure; it is homogeneous along the fibers, and inhomogeneous albeit periodic in the perpendicular direction. Such non-Euclidean plates have not been  studied previously, and one would tend at first to think that such surfaces are flat ``on average". It was therefore of a surprise when these monolayers were found to bend along the fibers. It is  one of the goals of the present paper  to resolve this pattern formation mechanism.

The first question of interest is why would a monolayer that is for all practical purposes homogeneous across the layer bend? We show that bending is the natural response to a uniaxially periodic non-Euclidean metric, i.e., it is a response to a metric incompatibility.  

Another characteristic of the experimental system studied in \cite{WMGTNSK12} is an almost-discontinuous reference metric. We show that as a result it is no longer possible to partition the range of parameters into ``thick sheets" and ``thin sheets"; one has to account for an interplay between the thickness of the sheet, the scale over which the reference metric varies between  bulk values, and the macroscopic dimensions of the stripes. In particular, we identify up to five distinct regimes of thickness, in each of which the the dependence of the configuration on the parameters is different. One of these regimes was not previously known, and its discovery sheds light onto the experimental and numerical results reported in \cite{KHHS12}.

These results show that  the homogenization of heterogenous materials is indeed highly non-trivial and scale dependent. Moreover, we expect that the multiplicity of energetic and conformational regimes could be even richer in systems in which the reference metric consists of several well-separated length scales.

\section{The model}

We model the composite sheets studied in \cite{WMGTNSK12} using the formalism of incompatible elasticity \cite{Wan66,Kro81}, and more specifically the dimensionally-reduced model of \emph{non-Euclidean plates} \cite{ESK08}.
A non-Euclidean plate is a two-dimensional surface, endowed with a Riemannian \emph{reference metric} $\ao$.  Other parameters of the plate are an elastic modulus $Y$, a Poisson ratio $\nu$, and a thickness $h$; all three parameters may be, in general, position dependent, however in this work we assume that they are spatially constant (note, however, that in the system studied in \cite{WMGTNSK12} the two gels have very distinct elastic moduli).

\begin{figure}[ht]
\begin{center}
\includegraphics[height=2.9in]{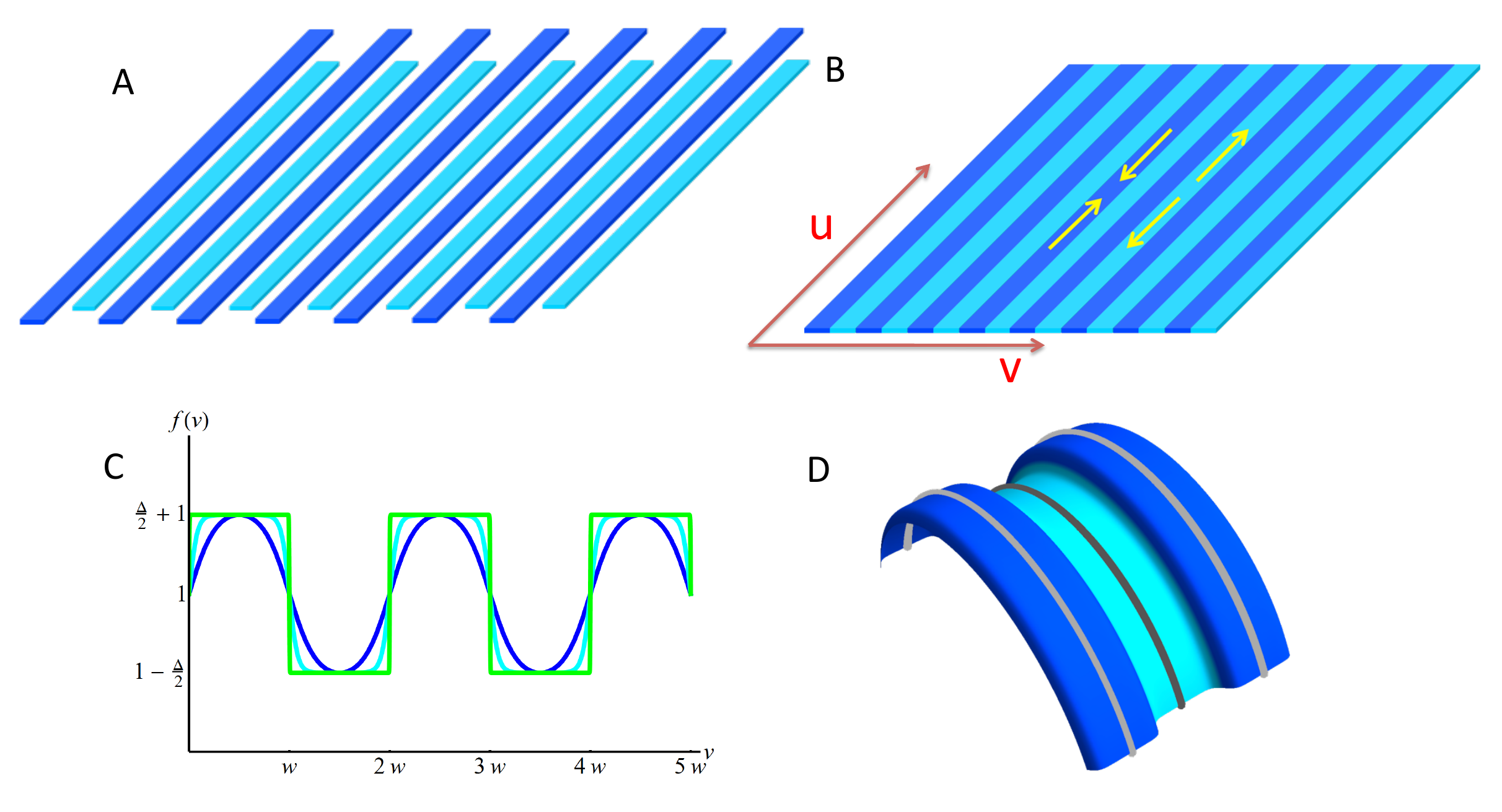}
\end{center}
\caption{Illustration of the experimental setting and observations in \cite{WMGTNSK12}. 
(A) Gel stripes that swell at different ratios upon stimulation arranged in an alternating array.  (B) A surface composed of alternating stripes. Upon stimulation internal stresses cause the respective compression and tension of the two types of gels (illustrated by yellow arrows). 
(C)  Functional dependence of the swelling profile on the $v$ coordinate. The three curves represent swelling profiles with different transition widths, $\lambda$. 
(D)  Cartoon of isometric embedding in the case of an abrupt metric transition. The sheet curves along the stripes while undulating in the perpendicular direction. The lines that are at the top of the undulations are longer than lines that are at the bottom of the undulations.}
\label{fig:1}
\end{figure}

We start by setting the reference metric $\ao$. 
For striped sheets, it is convenient to choose a system of coordinates with one of the axes parallel to the stripes. We denote points on the sheet by coordinates $(u,v)$, with the constant-$v$ parametric lines parallel to the stripes (see Figure~\ref{fig:1}B)  The symmetry of the system implies that the reference metric  depends only on the $v$-coordinate. Moreover, it is always possible to choose a so-called isothermal, or \emph{conformal} parametrization, such that the reference metric has the form
\beq
\ao(u,v) = f^2(v)
\brk{
\begin{array}{cc}
1 & 0 \\ 0 & 1 
\end{array}},
\label{eq:metric}
\eeq
where $f(v)$ is a local \emph{swelling profile} (shrinkage, if $f<1$).

For simplicity, we assume that prior to the stimulation both gels have the same width $w$, so that $f(v)$ is periodic with period $2w$. Inspired by the experimental setting, we assume that $f$ ranges between bulk values $1 \pm \Delta/2$, with a transition layer of width $\lambda \ll w$. Specifically, we choose the following  swelling profile (Figure~\ref{fig:1}C),
\beq
f(v) = 1 + \frac\Delta2  \, \frac{\tanh(\lambda^{-1} \sin(\frac{\pi v}{w}))}{\tanh \lambda^{-1} },
\label{eq:f}
\eeq
but all our results are insensitive to the precise functional form of $f(v)$.

Our goal is to study the equilibrium configuration of an unconstrained plate endowed with the metric \eqref{eq:metric}. We recall that any surface is uniquely determined (modulo rigid transformations) by its two fundamental forms: the first fundamental form, or \emph{metric} $\a$ and the second fundamental form, of \emph{curvature tensor} $\b$. In a non-Euclidean plate, the elastic energy associated with a configuration is the sum of two contributions: a \emph{stretching energy}, which penalizes for deviations of the metric $\a$ from the reference metric $\ao$, and a \emph{bending energy}, which penalizes for deviations of the curvature tensor $\b$ from a reference curvature tensor $\bo=0$ (non-zero reference curvature tensors distinguish plates from shells \cite{KS12}).

Naively, one would think that a state of zero energy can be attained by adopting a configuration in which $\a = \ao$ and $\b=\bo$. This is however not always possible, because the fundamental forms $\a$ and $\b$ are not independent; they are constrained by a set of differential relations---the Gauss-Mainardi-Codazzi equations \cite{DOC76}. Non-Euclidean plates and shells differ from Euclidean ones in that the reference forms $\ao$ and $\bo$ do not satisfy these constraints, and as a result, the equilibrium configuration is selected by a competitive interplay between both stretching and bending energies. 

The precise form of the stretching and bending energies are material dependent. For small metric strains and small curvatures, it is possible to approximate these energetic contributions by terms that are quadratic in both strain and curvature. Specifically, the total elastic energy $\En$ may be approximated by
\begin{equation}
\En = \int_{\Omega} \brk{ W_S + W_B } \, ds,
\label{FullEnergy}
\end{equation}
where $\Omega$ is the domain of parametrization, $W_S$ is the \emph{stretching energy density}, 
\begin{equation}
W_S = \frac{h}{8} \mathcal{A}^{\alpha\beta\gamma\delta} (a-\ao)_{\alpha\beta} (a-\ao)_{\gamma\delta},
\label{eq:WS}
\end{equation}
and $W_B$ is the \emph{bending energy density}, 
\begin{equation}
W_B = \frac{h^3}{24} \mathcal{A}^{\alpha\beta\gamma\delta} b_{\alpha\beta}b_{\gamma\delta}
\label{eq:WB}
\end{equation}
(as standard we adopt Einstein's summation convention). Here $\mathcal{A}$ is an \emph{elastic modulus tensor}, which for locally isotropic materials takes the form
\begin{equation}
\mathcal{A}^{\alpha\beta\gamma\delta}=\frac{Y}{1+\nu}
\brk{\frac{\nu}{1-\nu}\ao^{\alpha\beta}\ao^{\gamma\delta}+\ao^{\alpha\gamma}\ao^{\beta\delta}}.
\end{equation}
We take the domain of integration $\Omega$ to be
\[
\Omega = [0,\ell]\times[-w,w],
\]
with periodic boundary conditions in $v$; $\ell$ is the length of the stripes, which we assume to be longer than any other length scale, and
$ds$ denotes the differential area element,
\begin{equation}
ds = \sqrt{\det\ao} \, du dv.
\end{equation}

To summarize, the equilibrium configuration of the plate is postulated to be the ($v$-periodic) minimizer of the total elastic energy \eqref{FullEnergy}, where $W_S$ and $W_B$ are given by \eqref{eq:WS} and \eqref{eq:WB}, and the reference metric $\ao$ is given by \eqref{eq:metric} with swelling profile $f$ given by \eqref{eq:f}; the important parameters of $f$ are the \emph{swelling factor} $\Delta$, and the \emph{transition width} $\lambda$.

\section{Analysis}

\subsection{Why do thin striped plates buckle?}
\label{subsec:why}

Experiments show that thin striped plates of the type introduced in the previous section \emph{buckle}, i.e., they adopt a bent equilibrium configuration that breaks the plate's planar symmetry. In this section we explain in qualitative terms the mechanism that leads to this buckling. A more thorough analysis follows in the next subsections.

The prefactors in front of the stretching and bending energy densities \eqref{eq:WS},\eqref{eq:WB} have different dependences on the thickness of the plate $h$: the prefactor of the stretching content is linear in $h$ whereas the prefactor of bending content is cubic in $h$. For thick plates, the bending energy is dominant, and therefore the energy minimizing configuration is planar. For thin plates, the stretching energy is dominant, and  the plate will buckle, or bend, if it can thus lower its stretching energy. 

In a flat configuration the striped plates are metrically strained, because a flat configuration implies that both types of stripes have the same length, whereas their reference lengths are different (Figure~\ref{fig:1}A,B). Bending parallel to the stripes' longitudinal axis, thus turning the sheet into a cylindrical envelope, does not change this metric mismatch. If however we superimpose on top of a longitudinal bending a transversal undulation that is correlated with the stripes, then stripes that are at the top of the undulation are longer than stripes that are at the bottom (Figure~\ref{fig:1}D) and therefore the strain is reduced.

Such deformations cost bending energy, so that the equilibrium configuration has to be selected by an interplay between the stretching term, which as just described favors bent configurations, and the bending term, which favors planar ones. In \cite{WMGTNSK12}, a simplified semi-quantitative model was derived to predict shape selection in the extreme case of stiff fibers embedded in a soft swelling matrix. In the next section we perform a  quantitative analysis that reveals a fine balance between the stretching and the bending energies. We show that as the thickness of the sheet is reduced, the equilibrium configuration may be characterized by up to five distinct regimes, each characterized by a different dependence of the system's parameters.

\subsection{Surfaces of revolution}

We turn to study the minimization of the elastic energy \eqref{FullEnergy}, with a swelling profile \eqref{eq:f}. While a full solution of this problem is still beyond reach, a much simplified problem is obtained if restricting the space of configurations to \emph{surfaces of revolution}---an ansatz that is natural given that surfaces of revolution are consistently observed in experiments.  

For surfaces of revolution, both first and second fundamental forms are diagonal (in the chosen parametrization) and only depend  on the transverse coordinate $v$; we adopt the standard notation
\beq
\a = \brk{\begin{array}{cc}
E & 0 \\ 0 & G
\end{array}}
\qquad
\b = \brk{\begin{array}{cc}
L & 0 \\ 0 & N
\end{array}}.
\label{eq:forms}
\eeq
The four periodic functions $E(v)$, $G(v)$, $L(v)$, and $N(v)$ are constrained by the Gauss-Mainardi-Codazzi equations \cite{DOC76}, which given the imposed symmetries reduce to a pair of differential-algebraic equations,
\beq
L'=\frac{E'}{2 E}L+\frac{E'}{2 G}N
\qquad\qquad
L N=\frac{E'^2}{4 E} + \frac{E' G' }{4  G} - \frac{E''}{2}.
\label{eq:GMC}
\eeq
By straightforward integration methods we may obtain $L^2$ in terms $E$ and $G$,
\begin{equation}
L^2= c\, E-\frac{E'^2}{4 G}\,\,\,,
\label{GeneralSol}
\end{equation}
where $c$ is an integration constant. A direct substitution yields $N^2$ as well in terms of $E$ and $G$.

We have thus reduced the problem to that of minimizing the elastic energy $\En$ expressed as a functional of two $2w$-periodic functions, $E(v)$ and $G(v)$, and a single constant of integration $c$. Since the problem is symmetric in the longitudinal direction we may minimize the energy per unit length, which amounts to taking $\ell=1$; also we set for concreteness the Poisson ratio to be zero. Under these choices, the total energy measured in units of the stretching modulus $Yh/8$ takes the form,
\beq
\En = \int_{-w}^w \brk{(E-f^2)^2 + (G-f^2)^2}\, \frac{dv}{f^2} + \frac{h^2}{3} \int_{-w}^w \brk{L^2 + N^2}\, \frac{dv}{f^2}.
\label{eq:E2}
\eeq
The minimization of the energy \eqref{eq:E2} with respect to $E(v)$, $G(v)$, 
and $c$ can be  performed easily by numerical optimization methods.

\begin{figure}
\begin{center}
\includegraphics[height=2.6in]{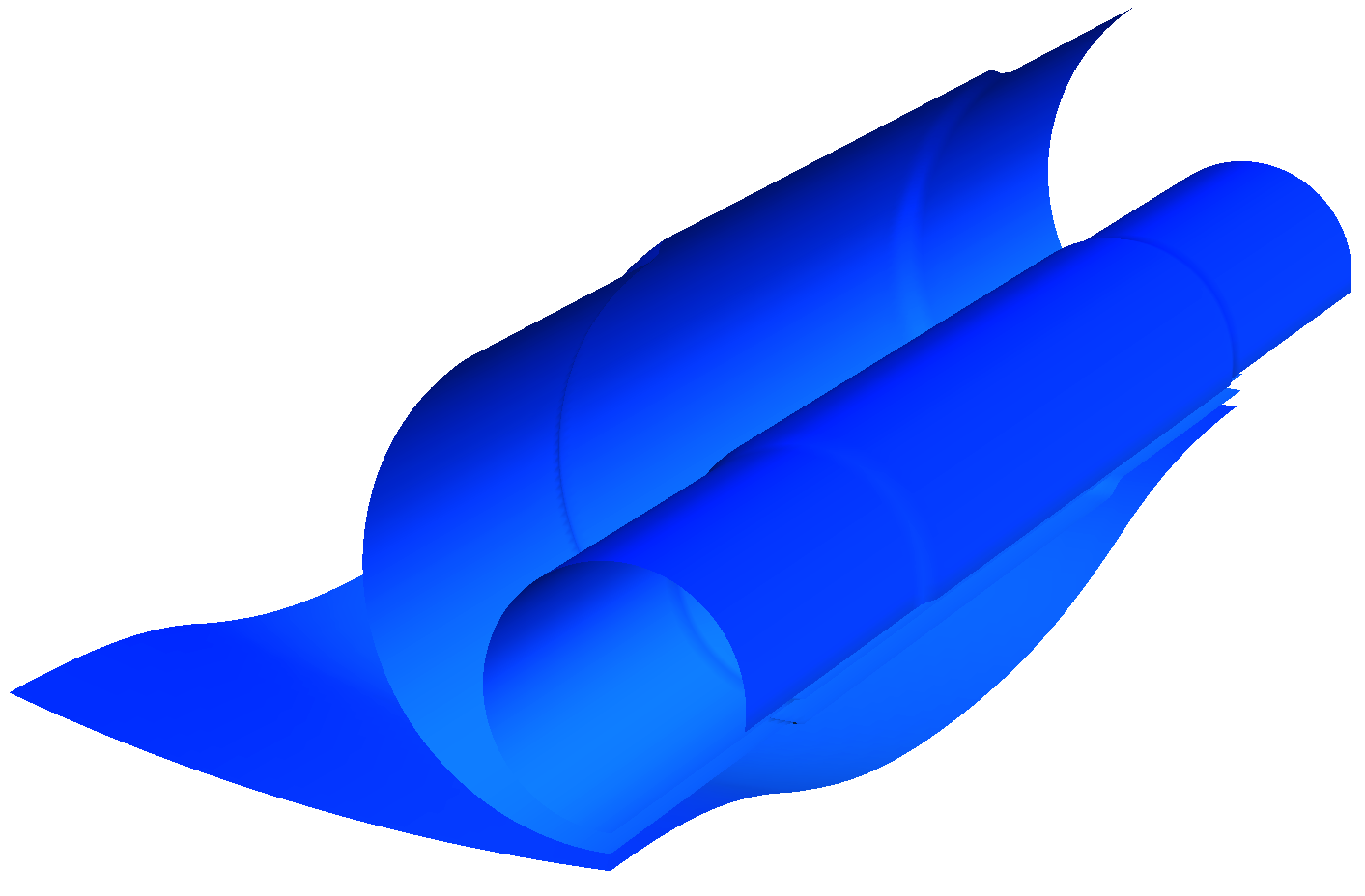}
\end{center}
\caption{Equilibrium configurations of one periodic unit for $\Delta=0.1$, $\lambda=1/160$, and three difference values of $h$. The smaller $h$, the larger is the longitudinal curvature, and the more noticeable is the metric transition.
}
\label{fig:1b}
\end{figure}

\subsection{Numerical resuts}

\begin{figure}
\begin{center}
\includegraphics[height=2.6in]{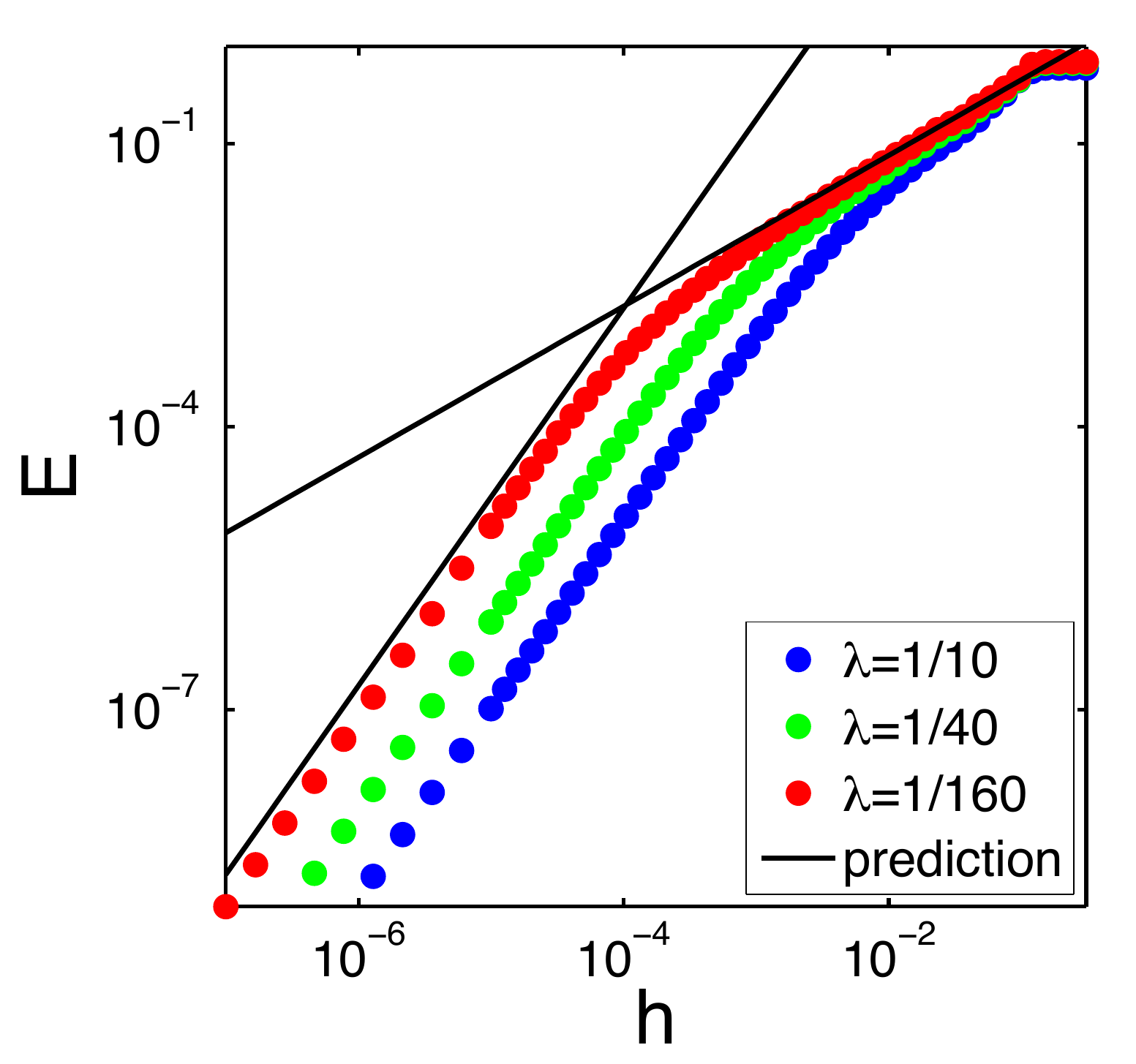}
\includegraphics[height=2.6in]{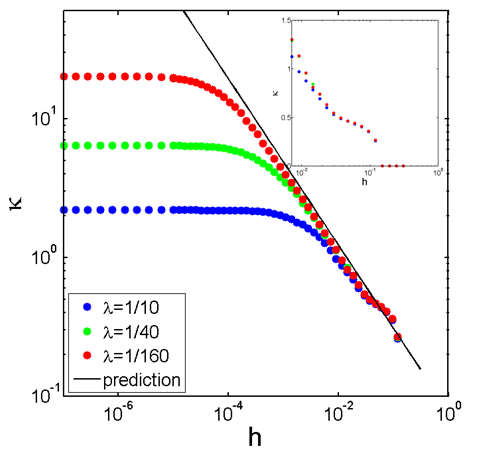}
\end{center}
\caption{(A) Total elastic energy  $\En$ (expressed in units of the stretching modulus $Yh/8$) versus  $h$ for $\Delta=0.1$ and three values of the transition width $\lambda$ (see legend). The solid curves correspond to the theoretical predictions $E\sim h^{2}$ and $E\sim h^{4/5}$.
(B) Mean longitudinal curvature $\kappa$ for the same parameters; the solid line corresponds to $\kappa\sim h^{-3/5}$. The inset shows that there is almost no dependence of $\kappa$ on $\lambda$ close to the buckling threshold.
}
\label{fig:2}
\end{figure}

\begin{figure}
\begin{center}
\includegraphics[height=2.6in]{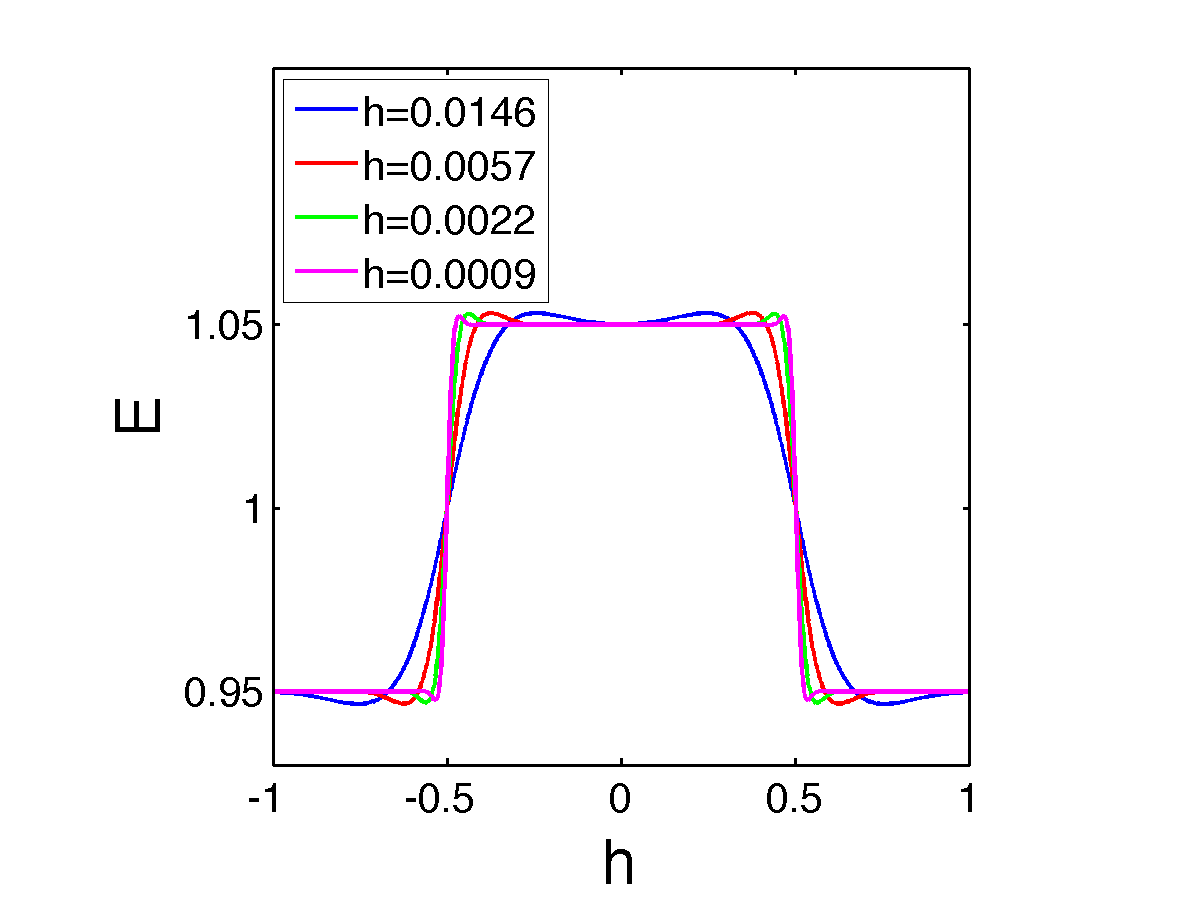}
\includegraphics[height=2.6in]{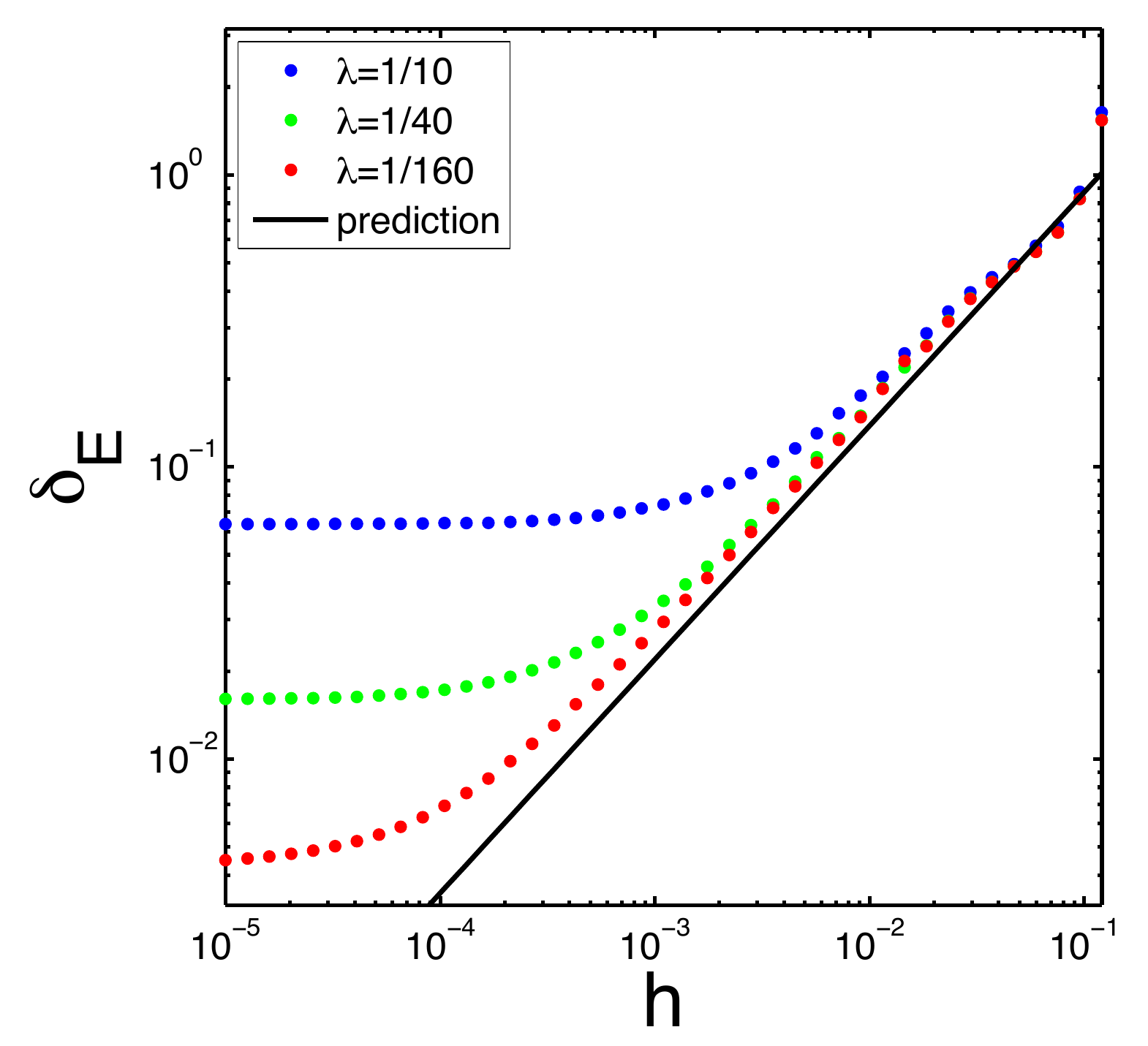}
\includegraphics[height=2.6in]{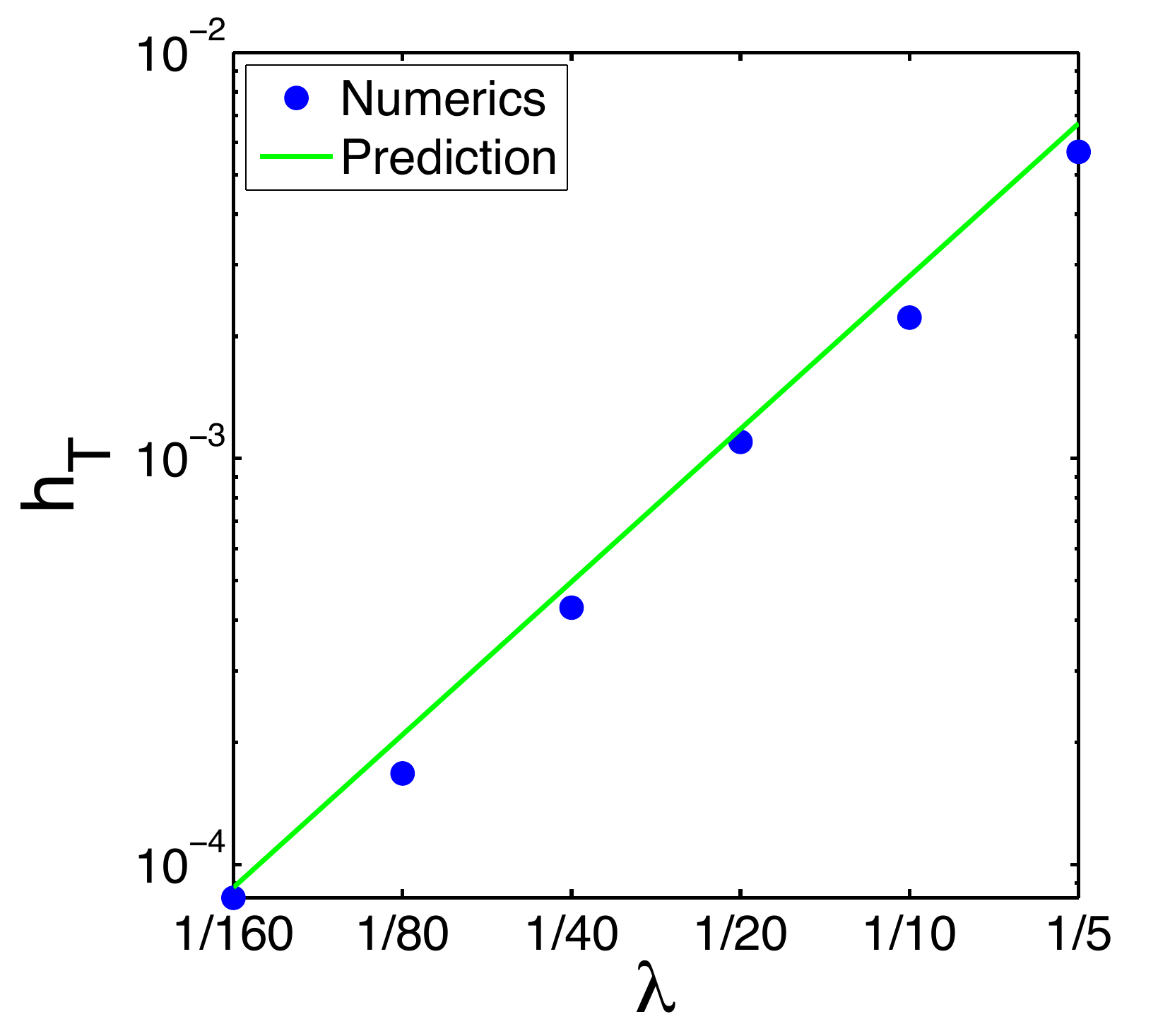}
\end{center}
\caption{(A) The metric component $E$ versus $v$ for $\Delta=0.1$, $\lambda=1/40$, and several values of $h$ (see legend). (B)
Log-log plot of an estimated value of $\delta_E$ versus $h$ for $\Delta=0.1$ and three different values of $\lambda$ (see legend); the solid line corresponds to the theoretical prediction $\delta_E\sim h^{4/5}$.
(C) An estimated value of the crossover thickness $h_T$ versus the transition width $\lambda$; the solid line correspond to the theoretical prediction $h_T\sim \lambda^{5/4}$.}
\label{fig:3}
\end{figure}

Computational results are shown in Figure~\ref{fig:1b}--\ref{fig:3}.
In Figure~\ref{fig:1b} we plot  equilibrium configurations for a choice of parameters of $\Delta=0.1$, $\lambda=1/160$, and three different values of $h$. These surfaces reveal two main features: (i) the smaller $h$ is, the larger is the longitudinal curvature; (ii) the smaller $h$ is, the more noticeable is the small scale of the swelling profile.  Both features reflect the approach of the actual metric $\a$ to the reference metric $\ao$.

Figure~\ref{fig:2}A displays a log-log plot of the elastic energy \eqref{eq:E2} as function of the thickness $h$ for $\Delta=0.1$, and  different values of the transition width $\lambda$ (the values are specified in the legend). 

Four distinct regimes are observed:

\begin{enumerate}
\item {\bfseries Planar configuration}:
For thick sheets we obtain, as expected, an energy that does not depend on $h$, which characterizes the \emph{plane-stress solution}. The buckling threshold $h_B$  is almost independent of the transition width $\lambda$.

\item{\bfseries Near threshold}:
The energy dependence on $h$ near the buckling threshold is almost independent of the transition width $\lambda$.

\item {\bfseries Isometric configuration}:
For $h\to0$, we have $\En\sim h^2$, which indicates that the equilibrium configuration converges to an exact isometry of the 2D reference metric $\ao$. Note however the strong dependence of the prefactor of $h^2$ on the transition width $\lambda$; the data indicates that  $\En \sim h^2/\lambda$.

\item {\bfseries Intermediate}:
For $h$ below  the buckling threshold $h_B$, but above a crossover values which we denote by  $h_T$  (see Figure~\ref{fig:2}A), a scaling of $\En \sim h^{4/5}$ is observed. Unlike the buckling threshold $h_B$, the crossover thickness $h_T$  strongly depends on $\lambda$. We note that an $h^{4/5}$ energy scaling has also been observed experimentally in \cite{KHHS12} for the case of a single-stepped metric, even though the theoretical prediction in this work was slightly different (we will return to this point in the Discussion).

\end{enumerate}

Figure~\ref{fig:2}B  
displays a log-log plot of a mean longitudinal curvature $\kappa$  as function of the thickness $h$ 
for the same parameters.
Here too we obtain four distinct regimes:

\begin{enumerate}
\item {\bfseries Planar configuration}:
For $h>h_B$ the equilibrium configuration is flat and therefore $\kappa=0$. 

\item {\bfseries Near threshold}:
Just below the buckling threshold $\kappa$ is almost independent of the transition width $\lambda$ (Figure~\ref{fig:2}B, inset). 

\item {\bfseries Isometric configuration}:
As $h\to0$ the curvature tends to a constant, as expected by the prediction that the equilibrium configuration converges to a bending minimizer among all isometries \cite{LP10,KS12}. The limiting curvature strongly depends on the transition width, $\kappa \sim 1/\lambda$.

\item {\bfseries Intermediate}:
In the intermediate regime, $h_T \ll h \ll h_B$, we obtain a scaling of $\kappa\sim h^{-3/5}$.

\end{enumerate}

Figure~\ref{fig:3}A displays the function $E(v)$  for $\Delta=0.1$, $\lambda=1/40$, and several values of $h$.  For small enough $h$, the value of $E$ far from the transition zone is close to the reference values, $1 \pm \Delta/2$. The transition between the two bulk values is over a distance which we denote by $\delta_E$, and which varies with $h$. As $h\to0$ the transition width $\delta_E$ converges to the transition width, $\lambda$, of the reference metric. 

In Figure~\ref{fig:3}B we plot an estimated value of the transition width $\delta_E$ versus $h$ for the same parameters as in Figure~\ref{fig:2}. The solid line corresponds to the theoretical prediction $\delta_E\sim h^{4/5}$. In Figure~\ref{fig:3}C we plot an estimated value of the crossover thickness $h_T$ versus the transition width $\lambda$. The solid line corresponds to theoretical prediction $h_T\sim h^{5/4}$.

\subsection{The isometric limit}

As $h\to0$ the equilibrium configurations converge to an isometric immersion of $\ao$ that minimizes the bending energy \cite{LP10,KS12}. Thus, the first fundamental form of the limiting configuration is given by the reference values,
\[
E(v) = G(v) = f^2(v).
\]
Substituting  $E$ and $G$ into \eqref{GeneralSol}, the second fundamental form is given by the one-parameter family of solutions:
\[
L =\sqrt{c \, f^2 - f'^2} 
\qquad\text{ and }\qquad
N = \frac{f'^2 - ff''}{\sqrt{c \, f^2 - f'^2}},
\]
and the corresponding bending energy is
\[
\EB = \frac{h^2}{3} \int_{-w}^{w} \BRK{c \, f^2 - f'^2
+  \frac{(f'^2 -ff'')^2}{c \, f^2 - f'^2}}  \frac{dv}{f^2}.
\]
The constant of integration $c$ is determined by a minimization of $\EB$. Differentiating $\EB$ with resect to $c$ we obtain the implicit equation:
\[
\int_{-w}^{w} \frac{(f'^2-f f'')^2}{(c \, f^2-f'^2)^2} \, dv = 2w.
\]
We have thus obtained a closed solution for the $h\to0$ limiting equilibrium configuration and the corresponding elastic energy.

The solution depends on the transition width $\lambda$ through its dependence on the swelling profile $f$.
Since $L^2>0$, it must be the case that for every $v$,
\[
c\, f^2(v) \ge f'^2(v),
\]
which yield a lower bound for $c$. 
Substituting \eqref{eq:f} and setting $v=0$, it follows that
\[
c \geq \frac{\Delta^2}{4\lambda^2}.
\]
For small enough $\lambda$ and $v$ in bulk regions we have $f \approx 1 + \Delta/2$ and $f'\approx 0$, in which case
\[
L \approx \sqrt{c\,(1 + \Delta/2)} \ge \frac{\Delta}{2\lambda},
\] 
which indicates that the longitudinal curvature should grow (at least) proportionally to $\lambda^{-1}$;
the data in Figure~\ref{fig:2}B shows that this is in fact the correct scaling.

\subsection{The intermediate asymptotic regime}

We now turn to analyze the energy scaling in the intermediate regime, where $h$ is below the buckling threshold $h_B$, but not too small (in a sense to be made precise).

As seen in numerical computations, in this intermediate regime the metric $\a$ is very close to the reference metric $\ao$ throughout most of the domain, except in a transition layer, which is larger than  $\lambda$. Moreover, the metric components $E$ and $G$ have different transition layers (data not shown), whose lengths we denote by $\delta_E$ and $\delta_G$, respectively; our computations show that $\delta_G\ll\delta_E$. Note that the composite sheet is mostly frustrated along the stripes, which means that the metric component $G$ can more easily  than to $E$ approach its reference value.
We will henceforth ignore the role of $G$ in the energy minimization; a more detailed analysis (not shown) taking $G$ into account reveals that this omission does not affect the results.

We  perform a scaling analysis to predict the dependence of the energy and the curvature on the various parameters. We start with the stretching energy $\ES$. The bulk  values of the diagonal entries of the reference metric are $1 \pm \Delta/2$, and the transition between these values is over a distance $\lambda$. For $\lambda\ll\delta_E$, the metric discrepancy in $E$ is of order $\Delta$  over a transition layer of order $\delta_E$, which yields a stretching energy of order
\beq
\ES \sim \Delta^2 \delta_E .
\label{eq:ES2}
\eeq

We now turn to estimate the bending energy $\EB$. Consider \eqref{GeneralSol}: since the derivatives of $E$ are small everywhere outside the transition layer, we have
\[
L^2 \approx c \qquad N \approx 0,  \qquad \text{ for $v$ not in transition zone},
\]
where we neglected $\Delta$ compared to $1$.
Inside the transition layer of size $\delta_E$, $E$ is of order one, whereas $E'\sim\Delta/\delta_E$ 
and $E''\sim\Delta/\delta_E^2$. 
Thus, inside the transition layer,
\beq
L^2 \sim  c - \frac{\Delta^2}{\delta_E^2}
\quad\text{ and }\quad
N^2 \sim  \frac{\Delta^2}{L^2 \delta_E^4}, 
\label{eq:LandN}
\eeq
and after straightforward algebraic manipulations the total bending energy (from both bulk and transition layer) satisfies the following asymptotic relation,
\begin{eqnarray*}
\EB &\sim 
h^2 w c + h^2 \delta_E \brk{c - \frac{\Delta^2}{\delta_E^2}} + 
h^2\delta_E \frac{\Delta^2}{(c - \Delta^2/\delta_E^2)\delta_E^4} \\
&\sim h^2 w  \frac{\Delta^2}{\delta_E^2} + h^2 (w+\delta_E) A  + 
h^2\delta_E \frac{\Delta^2}{A \delta_E^4},
\end{eqnarray*}
where $A = c - \Delta^2/\delta_E^2$. We then further neglect $\delta_E$ compared to $w$ in the middle term. 

The parameter $A$, which plays the same role as the integration constant $c$, does not affect the stretching energy, and therefore its value is determined by a minimization of the bending energy only. We get  $A  \sim \Delta w^{-1/2}  \delta_E^{-3/2}$, which upon substitution back in the bending energy yields
\beq
\EB \sim 
 \frac{h^2 \Delta w}{\delta_E^2}  \brk{\Delta + \sqrt{\frac{\delta_E}{w}}}.
\label{eq:EB2}
\eeq

We next turn to determine the value of the transition width $\delta_E$. Unlike $A$, $\delta_E$ affects both the stretching and the bending, and is therefore determined by a minimization of the total energy,
\[
\En \sim \Delta^2\delta_E + \frac{h^2 \Delta w}{\delta_E^2}  \brk{\Delta + \sqrt{\frac{\delta_E}{w}}}.
\]

It is unclear a priori  whether $\Delta$ dominates $\sqrt{\delta_E/w}$ or vice versa in the bracketed expression; both are small parameters. A straightforward asymptotic expansion reveals there are indeed two possible scalings for $\delta_E$,
\[
\delta_E \sim 
\cases{
w^{1/3} h^{2/3} & if $h/w \ll\Delta^3  \qquad\text{(Case 1)}$ \\
\Delta^{-2/5} w^{1/5} h^{4/5} & if $h/w \gg\Delta^3 \qquad\text{(Case 2)}$. 
}
\]
Case 1 corresponds to a regime in which the bending is dominated by $L^2$ in the bulk, whereas Case 2  corresponds to a regime where the bending is determined by a balance between $L^2$ in the bulk and  $N^2$  in the transition layer. 

These two asymptotic regimes yield the following  scalings for the energy,
\[
\En \sim \cases{
\Delta^2 w^{1/3} h^{2/3} & if $h/w \ll\Delta^3 \qquad\text{(Case 1)}$ \\
\Delta^{8/5}  w^{1/5}h^{4/5} & if $h/w \gg\Delta^3 \qquad\text{(Case 2)}$,
}
\]
and the following scalings for the longitudinal curvature,
\[
L \sim \cases{
\Delta^{1/2}w^{-1/2} h^{-1/2} & if $h/w \ll\Delta^3 \qquad\text{(Case 1)}$ \\
\Delta^{4/5} w^{-2/5} h^{-3/5}  & if $h/w \gg\Delta^3 \qquad\text{(Case 2)}$.
}
\]

A comparison to our computational results indicates that we obtained scaling laws that match Case 2. Indeed, $\Delta^3\sim 10^{-3}$, which means that for Case 1 to occur we need $h/w\ll 10^{-3}$, which is already close to the isometric regime. It takes simple algebra to see that three distinct regimes are obtained only if  $\lambda/w \ll \Delta^2$, in which case,
\[
\En \sim \cases{
\Delta^2 w \lambda^{-2} h^2 & if $h \ll w (\lambda/w)^{3/2} \qquad\text{(Isometric limit)}$ \\
\Delta^2 w^{1/3} h^{2/3} & if $w (\lambda/w)^{3/2} \ll h \ll w\Delta^3 \qquad\text{(Case 1)}$\\
\Delta^{8/5}  w^{1/5}h^{4/5} & if $w \Delta^3 \ll h \ll h_B \qquad\text{(Case 2)}$.
}
\]
If, however $\lambda/w \not\ll \Delta^2$, then we obtain only two distinct  regimes,
\[
\En \sim \cases{
\Delta^2 w \lambda^{-2} h^2 & if $h \ll h_T \qquad\text{(Isometric limit)}$ \\
\Delta^{8/5}  w^{1/5}h^{4/5} & if $h_T \ll h/w \ll h_B \qquad\text{(Case 2)}$,
}
\]
where
\[
h_T = \Delta^{1/2}w ֿ\, \brk{\frac{\lambda}{w}}^{5/4}.
\]
In Figure~\ref{fig:3}C we plot the dependence of $h_T$ on $\lambda$ and find that the predicted scaling is indeed correct.

\section{Discussion}

This paper presents a novel analysis of thin sheets of pre-stressed materials, in which the reference metric has a uniaxially periodic structure. Such a geometry is common in nature (e.g. in monolayers of fibrous tissue) as well as in man-made composite materials (e.g., \cite{WMGTNSK12}). As explained in Subsection~\ref{subsec:why}, such an incompatible metric favors a combination of longitudinal bending and transversal undulations. It should be noted, however, that the occurrence of other strain relieving mechanisms, such as wrinkles has not been ruled out.

Another characteristic of the reference metric is the abrupt transition between two locally Euclidean metrics. 
We showed that in such setting there is no simple distinction between a ``thick plate" and a ``thin plate". When varying the thickness of the sheet, the equilibrium conformation switches between several regimes, each satisfying its own scaling laws, thus reflecting different balances between energetic contributions. 

Our results are closely related to the theory and experiments reported in \cite{KHHS12}.
Rather than considering an array of alternating stripes, they considered a system which in our context corresponds to a single period. In addition, their experimental setting corresponds to a regime of parameters in which $\lambda\ll w$, i.e., a very abrupt metric transition, or equivalently, wide stripes. 
Their theoretic predictions  were scaling relations of $\En\sim h^{2/3}$ and $\kappa\sim h^{-1/2}$, which coincide with the regime we labeled Case 1. Yet, their numerical calculations clearly exhibit an energy scaling of $E\sim h^{4/5}$,  (Figure~2G in \cite{KHHS12}) which rather corresponds to our Case 2 regime. Thus, it appears as if their parameters correspond to the latter regime, which was only discovered in the present work.
It should be noted, however, that $\En\sim h^{2/3}$ is indeed the generic scaling for sufficiently small $\lambda$ and $h$.

The revelation that distinct scaling regimes may emerge with decreasing thickness is the most important contribution of this paper. We conjecture that it is a general feature in non-Euclidean plates in which the metric has a multi-scaled structure. At finite thickness $h$, bending dominates all features below a certain length scale, so that  metric discrepancies below this scale will remain unnoticed. Effectively, this is equivalent to the metric being smoothed at a cutoff scale determined by the thickness. As $h$ is further reduced, this cutoff scale decreases, and new features of the metric may thus manifest. 


\ack
RK was partially supported by the Israeli Science Foundation and by the Israel-US Binational Science Foundation. MM was supported  by the Israel-US Binational Science Foundation.
ES was partially supported by the ``Softgrowth" project of the European Research Council.

\section*{References}


\begin{thebibliography}{10}

\bibitem{WMGTNSK12}
Z.L. Wu, M.~Moshe, J.~Greener, H.~Therien-Aubin, Z.~Nie, E.~Sharon, and
  E.~Kumacheva.
\newblock Three-dimensional shape transformations of hydrogel sheets induced by
  small-scale modulation of internal stresses.
\newblock Submitted, 2012.

\bibitem{ESK08}
E.~Efrati, E.~Sharon, and R.~Kupferman.
\newblock Elastic theory of unconstrained {non-Euclidean} plates.
\newblock {\em J. Mech. Phys. Solids}, 57:762--775, 2009.

\bibitem{BG05}
M.~{Ben Amar} and A.~Goriely.
\newblock Growth and instabilities in elastic tissues.
\newblock {\em J. Mech. Phys. Solids}, 53:2284--2319, 2005.

\bibitem{Yav10}
A.~Yavari.
\newblock A geometric theory of growth mechanics.
\newblock {\em J. Nonlinear Sci.}, 20:781--830, 2010.

\bibitem{Wan67}
C.-C. Wang.
\newblock On the geometric structures of simple bodies, a mathematical
  foundation for the theory of continuous distributions of dislocations.
\newblock {\em Arch. Rat. Mech. Anal.}, 27:33--93, 1967.

\bibitem{Kro81}
E.~Kroner.
\newblock The physics of defects.
\newblock In R.~Balian, M.~Kleman, and J.-P. Poirier, editors, {\em Les Houches
  Summer School Proceedings}, Amsterdam, 1981. North-Holland.

\bibitem{KES07}
Y.~Klein, E.~Efrati, and E.~Sharon.
\newblock Shaping of elastic sheets by prescription of non-{Euclidean} metrics.
\newblock {\em Science}, 315:1116 -- 1120, 2007.

\bibitem{San09}
C.D. Santangelo.
\newblock Buckling thin disks and ribbons with {non-Euclidean} metrics.
\newblock {\em Europhys. Lett.}, 86:34003, 2009.

\bibitem{GV11}
J.~Gemmer and S.~Venkataramani.
\newblock Shape selection in {non-Euclidean} plates.
\newblock {\em Physica D}, 240:1536--1552., 2011.

\bibitem{DOC76}
M.P.~Do Carmo.
\newblock {\em Differential Geometry of Curves and Surfaces}.
\newblock Prentice Hall, 1976.

\bibitem{KHHS12}
J.~Kim, J.A. Hanna, R.C. Hayward, and C.D. Santangelo.
\newblock Thermally responsive rolling of thin gel strips with discrete
  variations in swelling.
\newblock {\em Soft Matter}, 8:2375--2381, 2012.

\bibitem{LM09}
H.~Liang and L.~Mahadevan.
\newblock The shape of a long leaf.
\newblock {\em Proc. Natl. Acad. Sci. USA}, 2009.

\bibitem{DB08}
J.~Dervaux and M.~{Ben Amar}.
\newblock Morphogenesis of growing soft tissues.
\newblock {\em Phys. Rev. Lett.}, 101:068101, 2008.

\bibitem{MP06}
M.~Marder and N.~Papanicolaou.
\newblock Geometry and elasticity of strips and flowers.
\newblock {\em J. Stat. Phys.}, 125:1069, 2006.

\bibitem{ESK11}
E.~Efrati, E.~Sharon, and R.~Kupferman.
\newblock Hyperbolic {non-Euclidean} elastic strips and minimal surfaces.
\newblock {\em Phys. Rev. E}, 83:046602, 2011.

\bibitem{FSDM05}
Y.~Forterre, J.M. Skotheim, J.~Dumais, and L.~Mahadevan.
\newblock How the {Venus} flytrap snaps.
\newblock {\em Nature}, 433:421--425, 2005.

\bibitem{DVR97}
C.~Dawson, J.F.V Vincent, and A.-M. Rocca.
\newblock How pine cones open.
\newblock {\em Nature}, 390:668, 1997.

\bibitem{AESK11}
S.~Armon, E.~Efrati, E.~Sharon, and R.~Kupferman.
\newblock Geometry and mechanics of chiral pod opening.
\newblock {\em Science}, 333:1726--1730, 2011.

\bibitem{KS12}
R.~Kupferman and J.P. Solomon.
\newblock A {Riemannian} approach to reduced plate, shell, and rod theories.
\newblock Submitted, 2012.

\bibitem{EZBF07}
R.~Elbaum, L.~Zaltzman, I.~Burgert, and P.~Fratzl.
\newblock The role of wheat awns in the seed dispersal unit.
\newblock {\em Science}, 316:884--886, 2007.

\bibitem{ATKDFRE11}
Y.~Abraham, C.~Tamburu, E.~Klein, J.W.C Dunlop, P.~Fratzl, U.~Raviv, and
  R.~Elbaum.
\newblock Tilted cellulose arrangement as a novel mechanism for hygroscopic
  coiling in the stork's bill awn.
\newblock {\em J. Roy. Soc. Interface}, 10.1098/rsif.2011.0395, 2011.

\bibitem{AAESK12}
H.~Aharoni, Y.~Abraham, R.~Elbaum, E.~Sharon, and R.~Kupferman.
\newblock Emergence of spontaneous twist and curvature in {non-Euclidean} rods:
  Application to erodium plant cells.
\newblock {\em Phys. Rev. Lett.}, 108:238106, 2012.

\bibitem{KHBSH12}
J.Kim, J.A. Hanna, M.~Byun, C.D. Santangelo, and R.C. Hayward.
\newblock Designing responsive buckled surfaces by halftone gel lithography.
\newblock {\em Science}, 335:1201--1205, 2012.

\bibitem{FZ55}
A.~Fahn and M.~Zohary.
\newblock On the pericarpial structure of the legumen, its evolution and
  relation to dehiscence.
\newblock {\em Phytomorph.}, 5:99--111, 1955.

\bibitem{FW72}
A.~Fahn and E.~Werker.
\newblock Anatomical mechanisms of seed dispersal.
\newblock In {\em Seed biology}. Academic Press, 1972.

\bibitem{RM09}
E.~Reyssat and L.~Mahadevan.
\newblock Hygromorphs: from pine cones to biomimetic bilayers.
\newblock {\em J. Roy. Soc. Interface}, 6:951--957, 2009.

\bibitem{Wan66}
C.-C. Wang.
\newblock On the stored-energy functions of hyperelastic materials.
\newblock {\em Arch. Rat. Mech. Anal.}, 23:1--14, 1966.

\bibitem{LP10}
M.~Lewicka and M.R. Pakzad.
\newblock Scaling laws for non-{Euclidean} plates and the {$W^{2,2}$} isometric
  immersions of {Riemannian} metrics.
\newblock {\em ESAIM: Control, Optimisation and Calculus of Variations},
  17:1158--1173, 2010.

\end{thebibliography}

\end{document}